# NUMERICAL SIMULATION OF THE ELECTRICALLY CONDUCTING FLOWS IN THE EXTERNAL MAGNETIC FIELD

*T. G. Elizarova[1], I. S. Kalachinskaya[2], Yu. V. Sheretov[3], I. A. Shirokov[2]*

[1] Institutre for the Mathematical Modeling, RAS,
[2] Moscow State University,
[3] Tver State University

E–mail adresses:  telizar@mail.ru, kalach@cs.msu.su, Yurii.Sheretov@tversu.ru, ivansh@ttk.ru

**Abstract**
The paper contains the original mathematical model that describes the incompressible electrically conducting fluid flows under the influence of the electromagnetic field – quasi-magneto-hydrodynamic (QMHD) equation system. Simplified inductionless approximation of QMHD system for the quasi-neutral liquid is constructed and implemented for numerical simulation of Marangoni convection in melted semiconductor suppressed by the static external magnetic field.

## 1. Introduction

Gravity and thermocapillary convective flow of the semiconductor melt can be suppressed by the external magnetic field. This effect is used to improve the semiconductor crystal characteristics. The possibility is studied widely experimentally [1, 2] and numerically [3, 4]. A numerical modeling of the interaction between the magnetic field and the electroconductive liquid seems to be a problem of great complexity.

A quasi-magneto-hydrodynamic (QMHD) model of the electrically conductive liquid flows, proposed in [5], is presented in this paper. QMHD system differs from the classical magneto-hydrodynamic (MHD) one by the additional diffusive terms. A simplified variant of the QMHD system for planar and axisymmetrical flows of the quasi-neutral semiconductor melt affected by the external magnetic field is constructed. To solve momentum and energy equations of QMHD system an explicit-time algorithm of the second order of the space accuracy is used. Poisson equation for the pressure is solved by the iterative numerical algorithm, convenient for the parallel implementation. The possibility of QMHD model is shown in the numerical simulations of the Marangoni convection in the cylindrical and rectangular cavities in a static magnetic field.

## 2. Mathematical model

We use the following notations: $r = const > 0$ for density, $\vec{u} = \vec{u}(\vec{x},t)$ for hydrodynamic velocity, $p = p(\vec{x},t)$ for dynamic pressure, $T = T(\vec{x},t)$ for temperature deviation from the average temperature $T_0 = const > 0$, $\vec{H} = \vec{H}(\vec{x},t)$ and $\vec{E} = \vec{E}(\vec{x},t)$ for magnetic and electric intensities respectively. The Oberbeck-Boussinesque approximation for the quasi-magneto-hydrodynamic (QMHD) system, proposed in [5], is used as the mathematical model. This approximation writes

$$\operatorname{div} \vec{u} = \operatorname{div} \vec{w}, \quad (1)$$

$$\frac{\partial \vec{u}}{\partial t} + \operatorname{div}(\vec{u} \otimes \vec{u}) + \frac{1}{r}\nabla p = -b\vec{g}T + \frac{1}{r}\operatorname{div} \Pi_{NS} + \operatorname{div}[(\vec{w} \otimes \vec{u}) + (\vec{u} \otimes \vec{w})] + \frac{1}{rc}[\vec{j}_e \times \vec{H}], \quad (2)$$

$$\frac{\partial T}{\partial t} + \operatorname{div}(\vec{u}T) = \operatorname{div}(\vec{w}T) + c\Delta T, \quad (3)$$

$$\operatorname{rot} \vec{E} = -\frac{1}{c}\frac{\partial \vec{H}}{\partial t}, \quad \operatorname{div} \vec{H} = 0, \quad \vec{j}_e = \frac{c}{4p}\operatorname{rot} \vec{H}. \quad (4)$$

Here

$$\Pi_{NS} = h[(\vec{\nabla} \otimes \vec{u}) + (\vec{\nabla} \otimes \vec{u})^T] \qquad (5)$$

is the Navier-Stokes shear stress tensor. Electric current density is given by

$$\vec{j}_e = s\left(\vec{E} + \frac{1}{c}[(\vec{u} - \vec{w}) \times \vec{H}]\right), \qquad (6)$$

where

$$\vec{w} = t\left[(\vec{u} \cdot \vec{\nabla})\vec{u} + \frac{1}{r}\vec{\nabla}p + b\vec{g}T - \frac{1}{rc}[\vec{j}_e \times \vec{H}]\right]. \qquad (7)$$

In (1)–(7), thermal expansion coefficient $b$, dynamic viscosity $h = rn$, thermal conductivity and thermal diffusivity $c$ and electric conductivity coefficient $s$ are positive constants. $\vec{g}$ is the gravity acceleration, $c$ is the velocity of light in vacuum. Relaxation time $t$ is given by $t = n/c_s^2$, where $n$ is kinematic viscosity coefficient and $c_s$ is sonic velocity in absence of the electromagnetic field. The QMHD equations differ from the classical MHD system by additional dissipative terms with the parameter $t$ having the dimension of time. As $t$ goes to zero, QMHD system reduces to the MHD one.

## 3. Problem formulation

Consider an axisymmetrical semiconductor melt flow in the cylindrical solid cavity. Computational domain is defined by $\Omega = \{(r,z): 0 < r < R, -A < z < A\}$. Uniform magnetic field $\vec{H}_0$ is directed along the cylinder axis, $\vec{g}$ is opposed to $\vec{H}_0$ (Fig. 1).

Inductionless approximation for (1)–(7) in cylindrical coordinates $(r,z)$ writes

$$\frac{1}{r}\frac{\partial}{\partial r}\left(r\frac{\partial p}{\partial r}\right) + \frac{\partial^2 p}{\partial z^2} = \frac{1}{t}\left[\frac{1}{r}\frac{\partial(ru_r)}{\partial r} + \frac{\partial u_z}{\partial z}\right] - \frac{1}{r}\frac{\partial}{\partial r}\left[r\left(u_r\frac{\partial u_r}{\partial r} + u_z\frac{\partial u_r}{\partial z} + Ha^2 u_r\right)\right] -$$
$$-\frac{\partial}{\partial z}\left(u_r\frac{\partial u_z}{\partial r} + u_z\frac{\partial u_z}{\partial z} - GrT\right), \quad (8)$$

$$\frac{\partial u_r}{\partial t} + \frac{1}{r}\frac{\partial(ru_r^2)}{\partial r} + \frac{\partial(u_z u_r)}{\partial z} + \frac{\partial p}{\partial r} = \frac{1}{r}\frac{\partial(r\Pi_{rr}^{NS})}{\partial r} + \frac{\partial\Pi_{zr}^{NS}}{\partial z} - \frac{\Pi_{jj}^{NS}}{r} +$$

$$+ \frac{2}{r}\frac{\partial(ru_r w_r)}{\partial r} + \frac{\partial(u_r w_z)}{\partial z} + \frac{\partial(u_z w_r)}{\partial z} - Ha^2(u_r - w_r), \quad (9)$$

$$\frac{\partial u_z}{\partial t} + \frac{1}{r}\frac{\partial(ru_r u_z)}{\partial r} + \frac{\partial(u_z^2)}{\partial z} + \frac{\partial p}{\partial z} = \frac{1}{r}\frac{\partial(r\Pi_{rz}^{NS})}{\partial r} + \frac{\partial\Pi_{zz}^{NS}}{\partial z} + \frac{1}{r}\frac{\partial(ru_z w_r)}{\partial r} +$$
$$+ \frac{1}{r}\frac{\partial(ru_r w_z)}{\partial r} + 2\frac{\partial(u_z w_z)}{\partial z} + GrT, \quad (10)$$

$$\frac{\partial T}{\partial t} + \frac{1}{r}\frac{\partial(ru_r T)}{\partial r} + \frac{\partial(u_z T)}{\partial z} = \frac{1}{r}\frac{\partial(rw_r T)}{\partial r} + \frac{\partial(w_z T)}{\partial z} + \frac{1}{\Pr}\left[\frac{1}{r}\frac{\partial}{\partial r}\left(r\frac{\partial T}{\partial r}\right) + \frac{\partial^2 T}{\partial z^2}\right]. \quad (11)$$

Here

$$w_r = t(u_r \frac{\partial u_r}{\partial r} + u_z \frac{\partial u_r}{\partial z} + \frac{\partial p}{\partial r} + Ha^2 u_r), \quad (12)$$

$$w_z = t(u_r \frac{\partial u_z}{\partial r} + u_z \frac{\partial u_z}{\partial z} + \frac{\partial p}{\partial z} - GrT). \quad (13)$$

Navier-Stokes shear stress tensor components are given by

$$\Pi_{rr}^{NS} = 2\frac{\partial u_r}{\partial r}, \quad \Pi_{zr}^{NS} = \Pi_{rz}^{NS} = \left(\frac{\partial u_r}{\partial z} + \frac{\partial u_z}{\partial r}\right), \quad \Pi_{jj}^{NS} = 2\frac{u_r}{r}, \quad \Pi_{zz}^{NS} = 2\frac{\partial u_z}{\partial z}.$$

In (8)–(13), the following assumptions are used:

- Electric intensity and electric current are neglected.
- Magnetic field is supposed to be uniform.

Equation (1) is rewritten identically in the form of Poisson equation (8).

System (8)–(13) is given in non-dimensional form. Parameters $r$, $z$, $t$, $u_r$, $u_z$, $w_r$, $w_z$, $p$, $T$ are scaled by $R$, $R$, $R^2/n$, $n/R$, $n/R$, $n/R$, $n/R$, $r(n/R)^2$, $\Theta$ respectively, where $\Theta$ stays for temperature at $(1,0)$. Dimensionless expression for $t$ writes

$$t = \frac{1}{Re_s^2}. \quad (14)$$

Grashof $Gr$, Hartman $Ha$, Prandtl $Pr$ and Reynolds $Re_s$ numbers, are introduced as follows:

$$Gr = \frac{gb\Theta R^3}{n^2}, \quad Ha = \frac{RH_0}{c}\sqrt{\frac{s}{h}}, \quad Pr = \frac{n}{c}, \quad Re_s = \frac{c_s R}{n}.$$

Assume the cylinder height $2A = 2R = 2$. Initial conditions are:

$$(u_r)|_{t=0} = (u_z)|_{t=0} = 0, \quad T|_{t=0} = 1 - |z|. \quad (15)$$

The boundary conditions, considering surface tension, write:

- symmetry axis ($r = 0$, $-1 < z < 1$):

$$u_r = 0, \quad \frac{\partial u_z}{\partial r} = 0, \quad \frac{\partial p}{\partial r} = 0, \quad \frac{\partial T}{\partial r} = 0; \quad (16)$$

- lateral boundary ($r = 1$, $-1 < z < 1$):

$$u_r = 0, \quad \frac{\partial u_z}{\partial r} = -\frac{Ma}{Pr}\frac{\partial T}{\partial z}, \quad \frac{\partial p}{\partial r} = 0, \quad T = 1 - |z|; \quad (17)$$

- lower ($0 < r < 1$, $z = -1$) and upper ($0 < r < 1$, $z = 1$) boundaries:

$$u_r = u_z = 0, \quad \frac{\partial p}{\partial z} = GrT, \quad T = 0. \quad (18)$$

Here

$$Ma = -\frac{\Theta R}{hc}\frac{\partial s_T}{\partial T}$$

is the Marangony number, $s_T$ stays for the surface tension coefficient. To eliminate the ambiguity in the pressure definition, the following condition was used: $p(0,0,t) = 0$. The problem formulation and the QMHD system for the planar flow of the melt are given in [6].

## 4. Numerical method and the computational results

The explicit finite-difference scheme of the second order in space, described in [6, 7], was used to solve (8)–(18). A steady solution was obtained as a limit at $t \to \infty$. Time stability criterion writes

$$\frac{1}{N_r N_z} \max \sum_{i,j} |(u^{up})_{ij} - u_{ij}| < e,$$

where $(u^{up})_{ij}$ is the velocity value at the upper time level, $e$ is the admissible error, $N_r$, $N_z$ are the numbers of the grid points in $r$ and $z$ directions. Poisson equation (8) was solved by means of the iterative multiprocessor-adapted method at each time step [8].

To increase numerical stability, $t$ (14) was re-defined by $t = 1/\text{Re}_s^2 + t_0$, where $t_0$ is the empirical regularizing parameter. In the case of the flows under consideration, the values of $\text{Re}_s$ are relatively great. Assuming $L = 1\ cm$, $n \sim 0.01\ cm^2/s$, $c_s \sim 10^5\ cm/s$, one has $\text{Re}_s \sim 10^7$. So we considered $t = t_0$.

The efficiency of the algorithm was examined by the numerical modeling. The grid convergence was tested also. The influence of the magnetic field on the intensity and structure of the convection of the melt in the cylindrical and square cavities (Fig. 1, 2) was investigated. In present work, we assume absence of the gravity ($Gr = 0$). The convection is driven by the surface tension only. The modeling was carried out for the following dimensionless parameters: $A = 1$; $Ma = 1000$; $\Pr = 0.018$; $Ha = 0, 50, 100$. The time step was $\Delta t = 10^{-7}$. Uniform special grids with square cells were used.

For the cylindrical cavity (Fig. 1) the dimensionless parameter $t = 2 \cdot 10^{-7}$, admissible error in the time stability criterion $e = 10^{-4}$. Computations were carried out on MVS-1000M multiprocessor machine using MPI standard. The number of time steps required for the stable solution decreased as the Hartman number increased (Table 1). This effect causes by the lower convection velocities in the magnetic field. Magnetic field also splits the convection into the multi-layered structure and makes the main vortex to move to the surface of the cavity.

Fig. 3–5 shows the stream function and the temperature contour levels. Stream function levels at the main vortex area and the temperature levels are equidistant. Stream function $y$ was calculated via

$$u_r - w_r = -\frac{1}{r}\frac{\partial y}{\partial z},\quad u_z - w_z = \frac{1}{r}\frac{\partial y}{\partial r},$$

supplemented by $y = 0$ at the boundary. Numerical simulation gives a symmetrical flow structure, because of the symmetrical distribution of the temperature at the lateral boundary and the no-gravity conditions ($Gr = 0$). Computations for $Ha = 100$ (Table 1) show the fast grid convergence of the numerical method.

In the case of the planar flow at the square cavity (Fig. 2), the adiabatic conditions for the temperature were used at the lower and upper boundaries. The condition T=1 was used at the left boundary, and T=0 was used at the right boundary. Dimensionless parameter $t = 2 \cdot 10^{-5}$. Admissible error in the time stability criterion was $e = 10^{-3}$. Fig. 6–10 and Table 2 present the results obtained. The arrows show the magnetic field intensity vector directions (marked as variants A and B).

Fig. 6–8 show the stream function and the temperature contour levels for the case of vertical magnetic field (variant A), $Ha = 0, 50$ and $100$, spatial grid $42 \times 42$. The contour levels are equidistant. Stream function $y$ was calculated via $u_x - w_x = \frac{\partial y}{\partial y}$, $u_y - w_y = -\frac{\partial y}{\partial x}$ with $y = 0$ at the boundary. As the magnetic field intensity increases, the convection velocity decreases, and the vortex moves to the free surface. In case of $Ha = 100$, the convection hardly distorts the temperature contour levels. Computations for $Ha = 50$ (Table 2) show the grid convergence of the numerical solution.

Fig. 9, 10 present the contour levels for the case of horizontal direction of the magnetic field intensity vector (variant B). Stream function levels at the upper area and the temperature levels are equidistant. The stream function absolute value peaks are greater then that in the case of vertical magnetic field. As for the case of cylindrical cavity, the magnetic field breaks the single vortex into three ($Ha = 50$) and four ($Ha = 100$) small vortexes.

Finite-difference algorithm based on classic MHD equations in the inductionless approximation was used in [4] to solve the similar problem. An implicit scheme of the third order spatial accuracy was constructed. To verify the accuracy of the new QMHD numerical method, authors carried out the melt flow simulation according with the conditions [4]. The simulation for $Ha = 50$ using the $42 \times 42$ spatial grid leaded to $y_{min} = -43.5$, where $y_{min}$ is the stream function minimum value at the main vortex center. Authors of [4] obtained $y_{min} = -44.2$. The flow structure was nearly identical in the both cases. The comparison demonstrates the good accuracy of the numerical method proposed in the present paper.

## 5. Conclusion

The original mathematical model to describe the incompressible electrically conducting fluid flows under the influence of the electromagnetic field – quasi-magneto-hydrodynamic (QMHD) equation system is presented. The simplification of QMHD system for inductionless quasi-neutral liquid flow in magnetic field is proposed. QMHD-based numerical algorithm (a time-explicit finite difference scheme with regularizing components) is constructed. This algorithm ensures the fast grid convergence and the stability of the numerical solution due to additional dissipation that differs QMHD and classical MHD system.

Numerical simulation of the Marangoni convection in the semiconductor melt flows, affected by the static external magnetic field, was carried out. It was established that magnetic field suppresses the convection and forces it to move to the free surface. Magnetic field with the parallel to the free surface intensity vector splits the convection into the multi-layered structure.

A comparison of the results obtained via the QMHD-based algorithm and the results obtained by well-developed numerical method based on classical MHD system ensures the accuracy of the relatively simple QMHD approach.

This work has been supported by grant INTAS No 2000-0617.

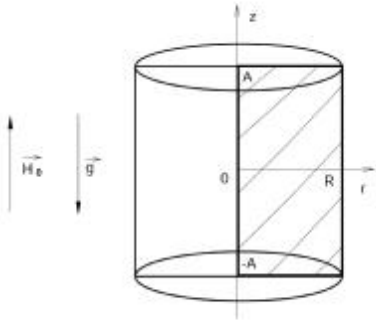

Fig. 1. Cylindrical cavity flow computational domain (hatched)

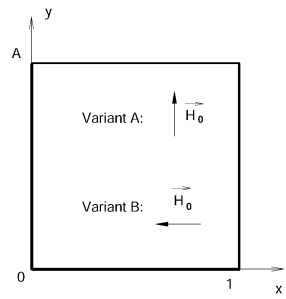

Fig.2. Square cavity flow computational domain

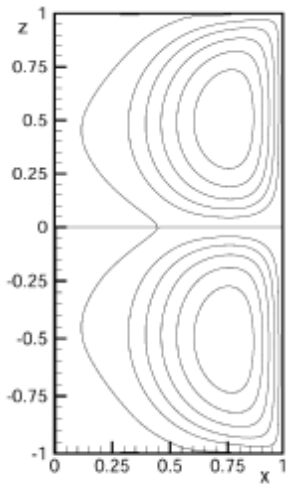

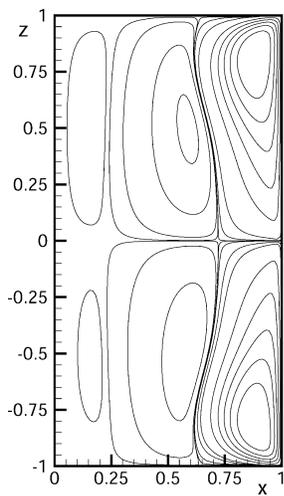

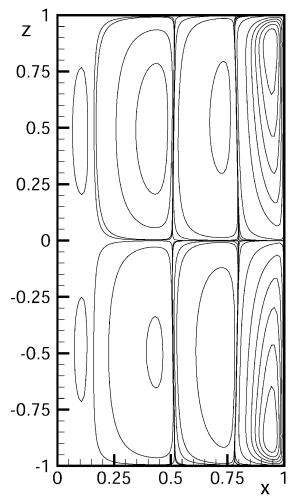

Fig. 3. Cylindrical cavity flow, Ha=0, stream function (above) and temperature (below) levels

Fig. 4. Cylindrical cavity flow, Ha=50, stream function (above) and temperature (below) levels

Fig. 5. Cylindrical cavity flow, Ha=100, stream function (above) and temperature (below) levels

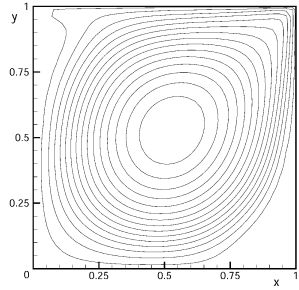
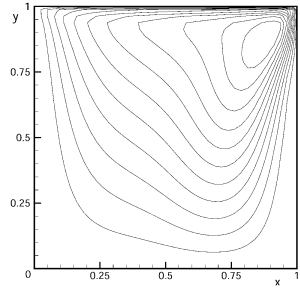
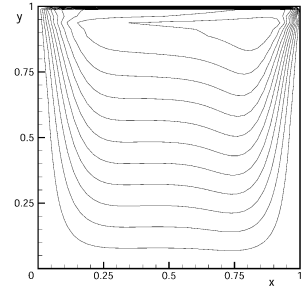
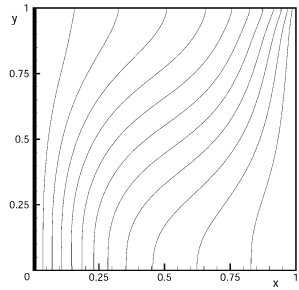
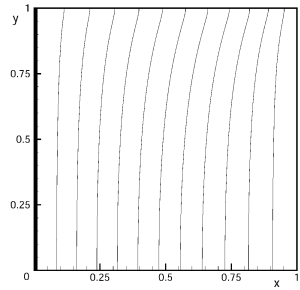
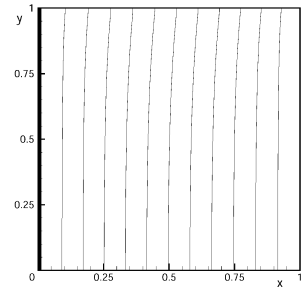

| Fig. 6. Square cavity flow, Ha=0, stream function (above) and temperature (below) levels | Fig. 7. Square cavity flow, Ha=50, vertical magnetic field (variant A), stream function (above) and temperature (below) levels | Fig. 8. Square cavity flow, Ha=100, vertical magnetic field (variant A), stream function (above) and temperature (below) levels |

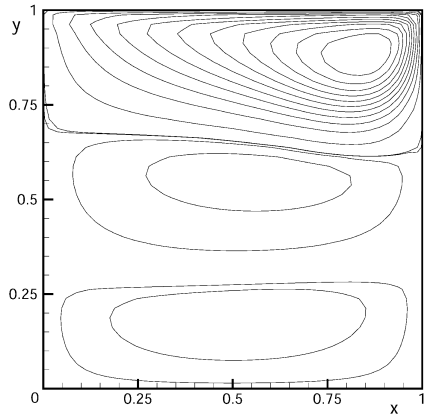
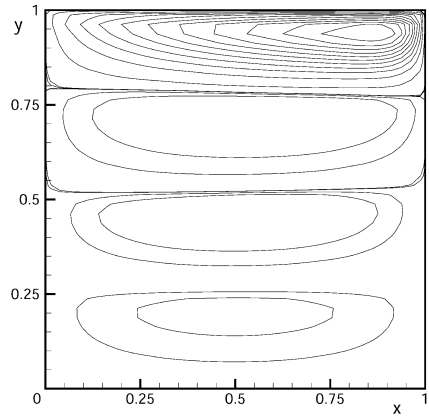
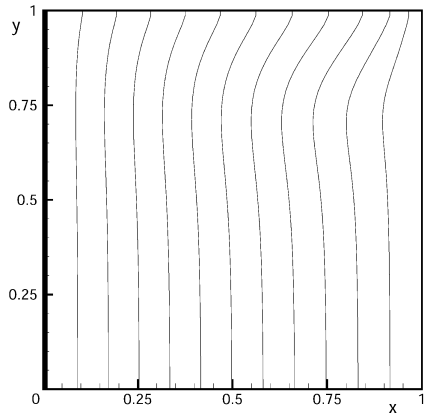
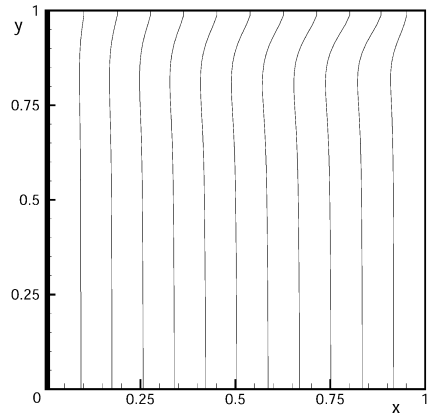

| | |
|---|---|
| Fig. 9. Square cavity flow, Ha=50, horizontal magnetic field (variant B), stream function (above) and temperature | Fig. 10. Square cavity flow, Ha=100, horizontal magnetic field (variant B), stream function (above) and temperature |

**Table 1. Cylindrical cavity flow modeling parameters**

| $Ha$ | Spatial grid, $N_r \times N_z$ | Number of time steps required for convergence | Stream function minimum $\psi_{min}$ |
|---|---|---|---|
| 0 | 82×162 | 569477 | –249.6 |
| 50 | 82×162 | 84326 | –52.2 |
| 100 | 82×162 | 45400 | –37.1 |
| 100 | 162×322 | 45574 | –37.0 |

**Table 2. Square cavity flow modeling parameters**

| $Ha$ | Spatial grid, $N_x \times N_y$ | Number of time steps required for convergence | Stream function minimum $\psi_{min}$ |
|---|---|---|---|
| 50 ↑ (var. A) | 22x22 | 437210 | –23.83 |
| 0 | 42x42 | 1500000 | –134.3 |
| 50 ↑ (var. A) | 42x42 | 440130 | –22.18 |
| 100 ↑ (var. A) | 42x42 | 353705 | –5.224 |
| 50 ← (var. B) | 42x42 | 441100 | –47.785 |
| 100 ← (var. B) | 42x42 | 375754 | –41.124 |
| 50 ↑ (var. A) | 82x82 | 439553 | –21.88 |